\documentclass[fleqn,usenatbib]{mnras}

\usepackage{newtxtext,newtxmath}

\usepackage[T1]{fontenc}
\usepackage{ae,aecompl}

\usepackage{graphicx}	
\usepackage{amsmath}	
\usepackage{amssymb}	

\title[Cosmic ray carrot]{Can a cosmic ray carrot explain the ionization level in diffuse molecular clouds?}

\author[Recchia et al.]{
S. Recchia,$^{1}$\thanks{E-mail: recchia@apc.in2p3.fr}
V. H. M. Phan,$^{1}$
S. Biswas$^{2}$
S. Gabici$^{1}$
\\
$^{1}$APC, AstroParticule et Cosmologie, Universit\'e Paris Diderot, CNRS/IN2P3, CEA/Irfu, Observatoire de Paris, Sorbonne Paris Cit\'e,\\10, rue Alice Domon et L\'eonie Duquet, 75205 Paris Cedex 13, France\\
$^{2}$ Raman Research Institute, C.V. Raman Avenue, Sadashivanagar, Bangalore 560080, India 
}

\date{Accepted XXX. Received YYY; in original form ZZZ}

\pubyear{2018}

\begin{document}
\label{firstpage}
\pagerange{\pageref{firstpage}--\pageref{lastpage}}
\maketitle

\begin{abstract}
Low energy cosmic rays are the major ionization agents of molecular clouds. 
However, it has been shown that, if the cosmic ray spectrum measured by Voyager 1 is representative of the whole Galaxy, the predicted ionization rate in diffuse clouds fails to reproduce data by $1-2$ orders of magnitude, implying that an additional source of ionization must exist. One of the solutions proposed to explain this discrepancy is based on the existence of an unknown low energy (in the range 1 keV-1 MeV, not probed by Voyager) cosmic ray component, called \textit{carrot} when first hypothesized by Reeves and collaborators in the seventies.
Here we investigate the energetic required by such scenario.
We show that the power needed to maintain such low energy component is comparable of even larger than that needed to explain the entire observed cosmic ray spectrum.  
Moreover, if the interstellar turbulent magnetic field has to sustain a \textit{carrot}, through second-order Fermi acceleration, the required turbulence level would be definitely too large compared to the one expected at the scale resonant with such low energy particles. 
Our study basically rules out all the plausible sources of a cosmic ray \textit{carrot}, thus making such hidden component unlikely to be an appealing and viable source of ionization in molecular clouds.
\end{abstract}

\begin{keywords}
clouds -- cosmic rays -- magnetic fields
\end{keywords}



\section{Introduction}
The ionization level of molecular clouds (MCs) is a crucial ingredient that determines their chemistry and the coupling between the gas and the magnetic field (see e.g \citealt{Dalgarno-2006} for a review).
The ionization rate of MCs is observed to decrease with increasing  cloud column density, with values that can reach $\rm \approx 10^{-15} s^{-1}$ in diffuse clouds, down to $\rm \approx 10^{-17} s^{-1}$ in denser clouds
(see \citealt{Padovani-2013-rev} and references therein).

Cosmic rays (CRs), especially the low energy ones (below $\rm \approx 1~GeV$), are widely recognized (see e.g \citealt{Padovani-2009}) as the major, or most likely the only, candidate able to ionize the interior of MCs, being the other main sources of ionization, namely UV photons and X-rays, unable to penetrate deeply inside MCs (\citealt{McKee-1989, Krolik-1983, Silk-1983}).

Locally, the interstellar low energy CR spectrum has been measured down to few MeV by Voyager 1, which is now thought to be far enough form the Sun, as to be unaffected by the solar modulation (\citealt{Stone-2013, Krimigis-2013, Cummings-2016}).

Several theoretical estimates of the CR induced ionization in MCs have been presented in the literature, starting form the pioneering works of \citet{Hayakawa-1961} and \citet{Spitzer-1968}, based on a simple extrapolation to low energies of the observed CR spectrum, to more refined models which take into account the propagation and energy losses of CRs in clouds (\citealt{Skilling-1976, Cesarsky-1978, Morfill-1982, Padovani-2009, Morlino-2015, Schlickeiser-2016, Ivlev-2018, Phan-2018}). 
In particular, \citet{Phan-2018} showed that if one assumes that the average low energy proton and electron spectrum in the Galaxy is the same as measured by Voyager 1, the inferred ionization rate inside diffuse MCs is $\rm \sim 1-2$ orders of magnitude smaller than the observed one. 
As pointed out by \citet{Phan-2018}, improvements of these models that, for instance, include also a description of dense and clumpy media and a more realistic modeling of the transition between different phases of the interstellar medium (ISM), are unlikely to enhance the predicted ionization rate by such large factor (see also \citealt{Morlino-2015}).

Thus, in order to reconcile predicted and measured ionization rates, one should either invoke a new source of ionization inside MCs, or question the validity of assuming the Voyager 1 spectrum to be representative of the whole CR spectrum in Galaxy.
Several possibilities have been put forward:
i) the possible presence of MeV CR accelerators inside MCs
(see e.g \citealt{Padovani-2015, Padovani-2016});
ii) inhomogeneities in the distribution of low energy CRs in the Galaxy (see e.g \citealt{Cesarsky-1975,Gabici-2015, Nobukawa-2015, Nobukawa-2018}); 
iii) the existence of a still unknown CR component emerging at energies below few MeV (the smallest energy detected by Voyager 1). Such component, called \textit{carrot}, was first proposed by \citet{Meneguzzi-1971} to explain the abundances of light elements, and has recently been reconsidered by \citet{Cummings-2016} (who called it {\it suprathermal tail}).\\

In this paper we focus on the \textit{carrot} scenario and we analyze in detail the implications of the possible presence of a CR population at energies below few MeV.
\footnote{We do not consider here the effect of the CR carrot on the production of light elements.
For a recent review of this topic see \citealt{Tatischeff-2018-review}}
In particular,  we estimate the power that has to be injected in low energy CRs in order to keep in the whole Galactic disk a population able to account for the observed ionization rate in MCs. We do so by assuming that the \textit{carrot} component is uniformly distributed both inside clouds and in the rest of the ISM. The power estimated in that way represents a very conservative lower limit, since in a more realistic scenario low energy CRs present in the ISM penetrate the cloud and their transport and energy losses in MCs have to be taken into account. As shown by \citet{Phan-2018}, the ionization rate predicted in this case would be smaller than in the simple scenario presented here. 
Here we show that, due to the relatively short ($\lesssim 10^5$ yr, see Eq.~\ref{eq:t-loss}) lifetime of sub MeV CRs in the ISM,  in order to maintain a very low energy and hidden CR component able to explain the observed ionization rates, it would be necessary for the potential sources to inject in the ISM a power  comparable to or larger than that needed to explain the whole observed CR spectrum. 
This result poses a serious concern on the viability of a \textit{carrot} scenario.

We also explore the implications of assuming that  such component be accelerated by the turbulent magnetic field in the ISM, through second-order Fermi acceleration (see e.g \citealt{Osborne-1998, Jokipii-2001, Drury-2014, Drury-2017}). However, we show that in this case the level of turbulence required at the scale resonant with CRs at the relevant energies is much larger than the one usually accepted. 
This brings additional support to the idea that a CR \textit{carrot} at energies below the smallest one detected by Voyager 1 fails to provide a solution to the problem of the ionization rate in MCs.
\section{Power requirement}
\label{sec:1}
Let us assume the presence of a CR (electron and/or proton) component at a given energy $\rm \Tilde{E} \lesssim 3~MeV$ (energies smaller than those detected by Voyager 1), uniformly distributed in the whole Galactic disk, including the interior of MCs. 
For simplicity, we assume that the distribution function of such component is:
\begin{equation}\label{eq:f-delta}
    f(E)= A\delta(E-\Tilde{E}),
\end{equation}
where $A$ is a normalization constant that has to be determined.

We do so by imposing that the $H_2$ ionization rate produced by CRs (electrons or protons) with the  distribution function given by Eq.~\ref{eq:f-delta}, equals the average value, $\rm \xi\approx 4\times10^{-16} s^{-1}$, detected in diffuse clouds (see e.g \citealt{Indriolo-2009}).

Such ionization rate can be computed, following the approach by \citet{Padovani-2009} and \citet{Phan-2018}, as
\begin{equation}\label{eq:xi-p}
         \xi^p=\int_{I}^{E_{Max}}f_p(E)v_p\left[\left( 1+\phi_p(E)\right)\sigma^p_{ion}(E) +\sigma_{ec}(E) \right]dE
    \end{equation}
    
     \begin{equation}\label{eq:xi-e}
         \xi^e=\int_{I}^{E_{Max}}f_e(E)v_e\left[ 1+\phi_e(E)\right]\sigma^e_{ion}(E) dE.
    \end{equation}
 Here $f_{p(e)}(E)$ is the CR proton(electron) distribution function, $v_{p(e)}$ is the incident CR velocity, $\sigma^{p(e)}_{ion}$ is ionization cross section and $\sigma_{ec}$ is the electron capture cross section, $\phi_{p(e)}$ are the average secondary ionizations per primary ionization \citep{Krause-2015}, $I=15.603$ eV is the $H_2$ ionization potential.

Once determined the overall normalization of the \textit{carrot} distribution function, the power needed in order to sustain such component in the whole Galactic disk can be estimated as
\begin{equation}\label{eq:P-req}
   P(\Tilde{E})=\frac{A(\Tilde{E})\Tilde{E}V_{disk}}{\tau_{loss}(\Tilde{E})}.
\end{equation}
Here $V_{disk}$ is the disk volume (radius $\rm R_d \sim 15~kpc$, height $\rm h_d\sim 300~pc$) and 


\begin{align}\label{eq:t-loss}
&\tau_{loss,p}(E)\approx 6\,E_{keV}^{4/3} \, yr &  \text{for E in  1 keV-1 MeV}\\
&\tau_{loss,e}(E)\approx 3\times 10^2 E_{keV}\, yr &  \text{for E in  1 keV-1 MeV} \nonumber
\end{align}
are the approximate expressions for the CR proton and electron energy loss time in the Galactic disk. Such energy losses are mainly due to ionization losses in the neutral phases of the ISM and Coulomb losses in the ionized phases of the ISM (see e.g \citealt{Schlickeiser-2016}).

The expressions of Eq.~\ref{eq:t-loss} are computed as 
\begin{equation}
    \tau_{loss(p,e)}(E)=\frac{1}{\sum_i r_{i(p,e)f_i}},
\end{equation}
where $r_i$ and $f_i$ are the loss rate and filling factor, respectively, for the different phases of the ISM. The ISM is approximated  as mainly constituted by three phases (see e.g \citealt{Osterbrock-1989}): 1) warm neutral medium (WNM), mostly made of neutral atomic hydrogen (density $\rm \approx 0.5\, cm^{-3}$, volume filling factor $\approx 25\%$, temperature $\approx 8000$ K); 2) warm ionized medium (WIM), mostly made of ionized atomic hydrogen (density $\rm \approx 0.5\, cm^{-3}$, volume filling factor $\approx 25\%$, temperature $\approx 8000$ K); 3) hot ionized medium (HIM), mostly made of ionized atomic hydrogen (density $\rm \approx 0.006\, cm^{-3}$, volume filling factor $\approx 50\%$, temperature $\approx 10^6$ K). 

\begin{figure}
	\centering
	\includegraphics[width=\columnwidth]{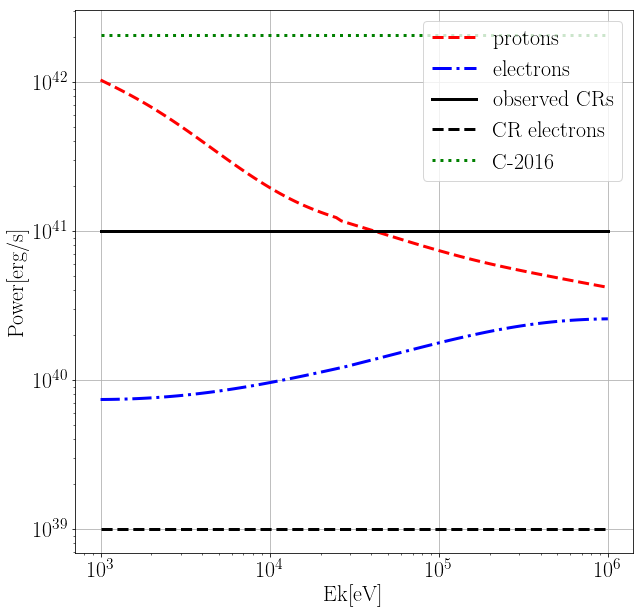}
    \caption{Power needed in CR   protons and electrons  in order to keep a \textit{carrot} at a given energy in the whole Galactic disk, able to predict (without taking into account the CR penetration in MCs) an ionization rate of $\rm 4\times 10^{-16}~s^{-1}$, as compared with the power needed to sustain the observed CR Galactic population (black, solid line) and the observed CR electron spectrum (black, dashed line), respectively. The line marked as C-2016 is the power required in CR protons in order to keep the suprathermal tail invoked in \protect\cite{Cummings-2016} in the whole Galactic disk .}
	\label{fig:1}
\end{figure}

\begin{figure}
	\centering
	\includegraphics[width=\columnwidth]{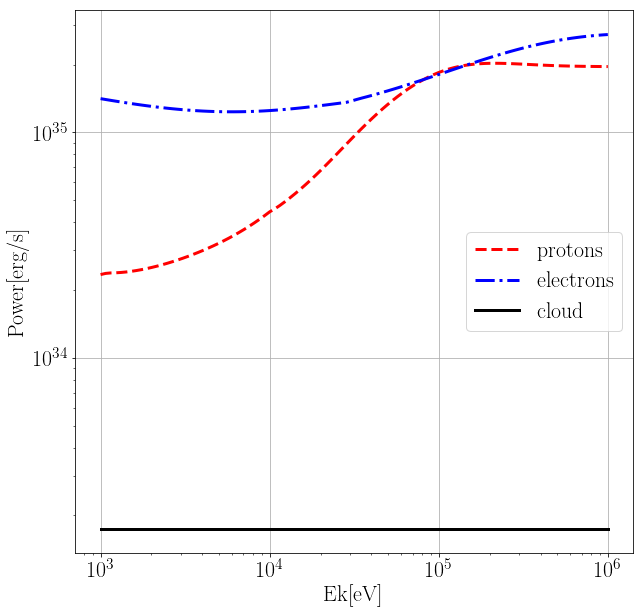}
    \caption{Power in CR electrons and protons  in order to keep a \textit{carrot} at a given energy within a cloud ($\rm n_c=100~cm^{-3}$, radius $R_c \sim 10$ pc), able to predict  an ionization rate of $\rm 4\times 10^{-16}~s^{-1}$. This is compared with the maximum power that a cloud can provide (black, solid line), given by $P_c = E_{grav}/\tau_{life}$ ($\tau_{life}\sim 10^7$ yr, $E_{grav}= \frac{3}{5}\frac{GM_c^2}{R_c}$).}
	\label{fig:2}
\end{figure}


In Fig.~\ref{fig:1} we show the power estimated in Eq.~\ref{eq:P-req} for CR electron and proton energies in the range $\rm 1~KeV-1~MeV$.  
We compare it with the total power (see e.g \citealt{Strong-2010}) injected by sources in the observed CR spectrum ($\approx 10^{41}$ erg/s) and electron spectrum ($\approx 10^{39}$ erg/s). 
We also show an estimate of the total power in CR protons needed to keep in the whole Galactic disk the suprathermal  tail invoked in \cite{Cummings-2016} as
\begin{equation}
    P_{C-2016}=\int_{1 KeV}^{1 MeV} \frac{4\pi J(E)E}{v_p(E)} \frac{V_{disk}}{\tau_{loss,p}(E)}dE \approx 2\times 10^{42} erg/s,
\end{equation}
where $J(E)$ is the CR proton flux of the suprathermal  tail (see Fig.~16 of \citealt{Cummings-2016}).

Remarkably, the plot in Fig.~\ref{fig:1} illustrates that, due to the short  lifetime of low energy CRs in the ISM (see Eq.~\ref{eq:t-loss}),  a CR \textit{carrot} (or the suprathermal tail of \citealt{Cummings-2016}) would require a power injection comparable or even larger than that already needed in order to account for the whole observed CR spectrum ($\approx 10^{41}$ erg/s). 
The situation  is especially dramatic dramatic for electrons, given that the observed CR power for them is $\approx 10^{39}$ erg/s.

Note that
$10^{41}$ erg/s roughly corresponds to 10\% of the total power of galactic supernova explosions. Since supernova remnants  are considered the major source of Galactic CRs (see e.g \citealt{Blasi-2013-Review}), our result implies that the existence of a CR \textit{carrot} would require either an
unreasonably large (in some cases even larger than 100 \%) CR acceleration efficiency for known CR sources, either the existence of another, much more powerful (and thus implausible), class of sources.

Notice that this result is not expected to change  with different assumptions on the spectral shape of the low energy component. In fact, the  required power injection is minimum for a proton (electron) \textit{carrot} at 1 MeV(1 keV), as shown in Fig.~\ref{fig:1}.  Any choice of a broader spectrum in the range 1keV-1MeV, able to predict the same ionization level in MCs, will inevitably imply a larger power injection. 

Moreover, this estimated power is a very conservative lower limit. In fact here we assumed that the unknown CR component is uniformly distributed in the whole Galactic disk and inside clouds. 
However, CRs have to penetrate the cloud.
As illustrated by \citet{Phan-2018}, taking into account this effect leads to a lower predicted level of ionization. This can be easily seen if, for instance, we consider the average distance travelled by CR electrons and protons inside a cloud before losing all their energy due to ionization losses, that we estimate as:
\begin{equation}
    L_{loss}(E)=v(E)\tau_{loss}(E, n_c),
\end{equation}
where
\begin{align}\label{eq:t-loss-cloud}
&\tau_{loss,p}(E, n_c)\approx
  \begin{cases}
    \frac{500}{n_{c}} yr & \text{for E in 1 keV-0.1 MeV}  \\
    1.1\times 10^4\frac{E_{MeV}^{4/3}}{n_{c}} yr & \text{for E in  0.1-1 MeV} 
  \end{cases}\\
&\tau_{loss,e}(E, n_c)\approx 10^5 \frac{E_{MeV}}{n_{c}} yr \qquad \quad \text{for E in  1 keV-1 MeV} \nonumber
\end{align}
are approximate expressions for the CR ionization loss time (\citealt{Padovani-2009, Phan-2018}) for CR electrons and protons  in a cloud of $H_2$ density given by $n_c$.

\begin{figure}
	\centering
	\includegraphics[width=\columnwidth]{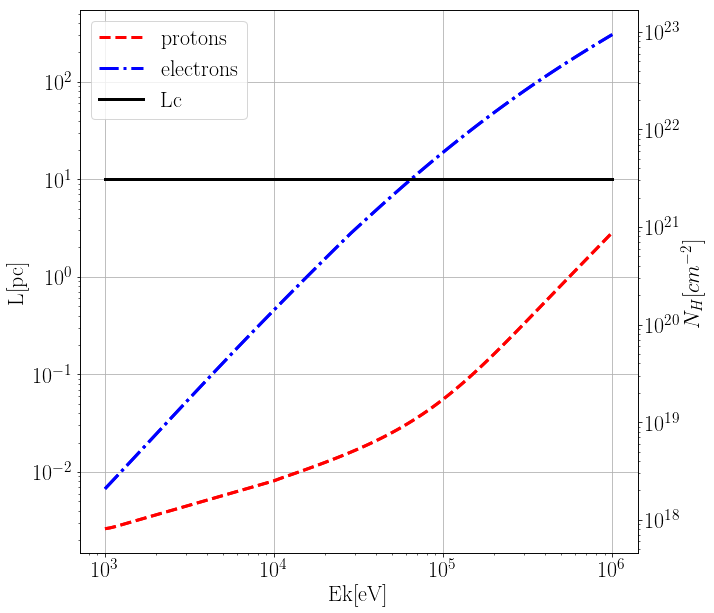}
    \caption{Average distance travelled by CR electrons (blue, dot-dashed line) and protons (red, dashed line) within a cloud ($\rm n_c=100~cm^{-3}$) in a loss time. The typical cloud size is assumed to be $L_c=10$ pc (black, solid line).}
	\label{fig:3}
\end{figure}

In Fig.~\ref{fig:3} we compare this typical distance for CR electrons and protons of energy in the range $\rm 1~KeV-1~MeV$ inside a cloud of $n_c=\rm 100~cm^{-3}$, with a typical cloud size $L_c = 10$ pc. The result is that protons of these energies  and electrons of $E \lesssim 0.1$ MeV would 
not even be able to cross a typical cloud. 

Notice  that also keeping a CR \textit{carrot} inside clouds instead that in the whole Galactic disk would lead an unsustainable power requirement. In this case, the rate at which CRs should be provided to the cloud can  be derived by using Eq.~\ref{eq:P-req} and Eq.~\ref{eq:t-loss-cloud},
provided that $V_{disk}$ is substituted with $V_{cloud}$.
The CR power obtained in this way is compared in Fig.~\ref{fig:2} to a characteristic maximal cloud power $P_c$ obtained by dividing the cloud gravitational energy $E_g=\frac{3}{5}\frac{GM_c^2}{R_c}$ by its typical lifetime $\tau_{life}$. 
We adopt typical cloud parameters $\rm R_c= 10~pc$, $\rm n_c= 100~cm^{-3}$ and $\rm \tau_{life}\sim 10^7~yr$ (see e.g \citealt{Heyer-2015}). 
The CR power largely exceeds $P_c$, making the \textit{carrot} scenario nonviable.

\section{Acceleration in the turbulent magnetic field}
\label{sec:2}
The results of Sec.~\ref{sec:1} already poses  serious doubts on the \textit{carrot} scenario  for the explanation of the observed ionization rate in MCs.

In order to bring additional support to this result, we also explore 
a possible major source of low energy CRs, namely the second order Fermi acceleration in the turbulent interstellar  magnetic field (see e.g \citealt{Osborne-1998, Jokipii-2001, Drury-2014, Drury-2017}).
The acceleration time-scale due to this process is given by (see Eq.~20 of \citealt{Drury-2014})
\begin{equation}\label{eq:t-acc}
    \tau_{acc}(E)=\frac{9}{4}\frac{D(E)}{v_A^2},
\end{equation}
where $D(E)=\frac{1}{3}\frac{v(E)r_L(E)}{I(k_{res})}$ is the spatial diffusion coefficient for particles of energy $E$ and $v_A=B_0/\sqrt{4\pi\rho}$ is the Alfv\'en speed. 
Here $v$ and $r_L$ are the particle velocity and Larmor radius, $I(k_{res})=W(k_{res})k_{res}$ is the level of turbulence, $(\delta B/B_0)^2$, at the resonant scale $k_{res}(E)=1/r_L(E)$, $B_0$ is the background magnetic field and $\rho$ the average mass density of the background medium.

\begin{figure}
	\centering
	\includegraphics[width=\columnwidth]{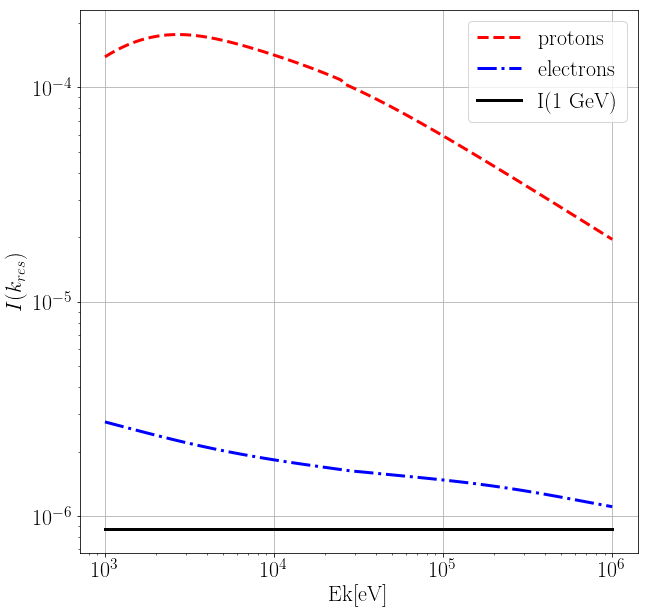}
    \caption{Level of magnetic turbulence needed to steadily maintain, through  second order Fermi acceleration,  sub MeV CR electrons and protons in the ISM ($B_0=3\mu$G). We show for comparison the turbulence expected at the scale resonant with $\sim 1$ GeV in order to account for accepted  values of the spatial diffusion coefficient ($\rm D(1~GeV)\sim  10^{28} cm^2/s$) at that energy.}
	\label{fig:4}
\end{figure}

Since low energy CRs lose energy in the ISM  on a time scale given by Eq.~\ref{eq:t-loss}, in order to keep a CR carrot at energy $E$ the level of magnetic turbulence at the resonant scale $k_{res}=1/r_L(E)$ have to be such that
\begin{equation}
    \tau_{loss}(E)= \tau_{acc}(E),
\end{equation}
namely 
\begin{equation}
    I(k_{res})=\frac{9}{4}\frac{v~r_L}{3v_A^2}\frac{1}{\tau_{loss}}.
\end{equation}
A plot of the needed level of  turbulence is shown in Fig.~\ref{fig:4} for $B_0=3\mu$G, in the case of CR electrons and protons of energy in the range $\rm 1~KeV-1~MeV$. 

Remarkably, the inferred $I(k_{res})$ at the energies relevant for this paper are  larger than, for instance, that expected at the scale resonant with $\sim 1$ GeV in order to account for  accepted (see e.g \citealt{Galprop-2011-propa}) values of the spatial diffusion coefficient ($\rm D(1~GeV)\sim 10^{28} cm^2/s$, $\rm I(1~GeV)\sim 9\times 10^{-7}$). This is quite unlikely to happen, since in any physical model of interstellar magnetic turbulence $I(k)$ is a decreasing function of $k$ (see e.g \citealt{Sridhar-1994, Goldreich-1995}).  
The present result, 
together with the results of Sec.~\ref{sec:1}, makes very difficult for a CR \textit{carrot} (or suprathermal tail) to represent a feasible  model able to reconcile the predicted and  observed ionization rates in MCs.

\section{Conclusions}
\cite{Phan-2018} showed that, if the CR electron and proton spectra measured by Voyager 1 are representative of the whole Galaxy, the penetration of such CRs inside diffuse MCs cannot account for the observed level of ionization in such clouds by $1-2$ orders of magnitude. This is an intriguing result that currently lacks an explanation. Among the solution proposed to this puzzle, there is the possibility that the CR electron and proton spectra may contain a still unknown component, called \textit{carrot}, at energies lower than the one detected by Voyager 1. 

In this paper we investigated this possibility, focusing in particular on the energetics involved if such a  \textit{carrot} has to account for the average ionization rate detected in diffuse MCs. 

We found that, due to the  energy losses suffered by low energy CRs in the ISM, the   power needed to be injected by the potential sources in such component is comparable or larger than that needed to explain the observed CR spectrum, even without taking into account the actual penetration of these low energy CRs inside clouds, which would make this energy requirement even more severe.

Moreover, if we consider the interstellar turbulent magnetic field as a possible source of this \textit{carrot}, through second-order Fermi acceleration, the required turbulence level would be definitely  too large compared to the one expected at the scale resonant with such low energy particles. 

Our study basically rules out, on an energy basis, any possible source of a CR \textit{carrot}, thus making such hidden component unlikely to be an appealing and viable source of ionization in MCs.

This conclusion encourages further studies of the possible solutions to the discrepancy between predicted and observed ionization rates in MCs.   
Among them, some promising ones remain the one already mentioned in the introduction and in \cite{Phan-2018}: i) the possible presence of sub-GeV CR accelerators inside MCs;
ii) prominent inhomogeneities  in the distribution of low energy CRs in the Galaxy. With this respect, we note that, given our peculiar location inside an ISM cavity (the local bubble, see e.g. \citealt{1998LNP...506..121C}), the CR spectrum measured by Voyager 1 might simply reflect local properties, rather than representing the typical spectrum of CRs in the Galaxy.

\section*{Acknowledgements}
SR and SG acknowledge support from the region \^{I}le-de-France under the DIM-ACAV programme, from the Agence Nationale de la Recherche (grant ANR- 17-CE31-0014), and from the Observatory of Paris (Action F\'ed\'eratrice CTA).
This project has also received funding from the European Union's Horizon 2020 research and innovation programme under the Marie Sklodowska-Curie grant agreement No 665850.




\bibliographystyle{mnras}
\bibliography{biblio} 

\begin{thebibliography}{}
\makeatletter
\relax
\def\mn@urlcharsother{\let\do\@makeother \do\$\do\&\do\#\do\^\do\_\do\%\do\~}
\def\mn@doi{\begingroup\mn@urlcharsother \@ifnextchar [ {\mn@doi@}
  {\mn@doi@[]}}
\def\mn@doi@[#1]#2{\def\@tempa{#1}\ifx\@tempa\@empty \href
  {http://dx.doi.org/#2} {doi:#2}\else \href {http://dx.doi.org/#2} {#1}\fi
  \endgroup}
\def\mn@eprint#1#2{\mn@eprint@#1:#2::\@nil}
\def\mn@eprint@arXiv#1{\href {http://arxiv.org/abs/#1} {{\tt arXiv:#1}}}
\def\mn@eprint@dblp#1{\href {http://dblp.uni-trier.de/rec/bibtex/#1.xml}
  {dblp:#1}}
\def\mn@eprint@#1:#2:#3:#4\@nil{\def\@tempa {#1}\def\@tempb {#2}\def\@tempc
  {#3}\ifx \@tempc \@empty \let \@tempc \@tempb \let \@tempb \@tempa \fi \ifx
  \@tempb \@empty \def\@tempb {arXiv}\fi \@ifundefined
  {mn@eprint@\@tempb}{\@tempb:\@tempc}{\expandafter \expandafter \csname
  mn@eprint@\@tempb\endcsname \expandafter{\@tempc}}}

\bibitem[\protect\citeauthoryear{{Blasi}}{{Blasi}}{2013}]{Blasi-2013-Review}
{Blasi} P.,  2013, \mn@doi [\aapr] {10.1007/s00159-013-0070-7}, \href
  {http://adsabs.harvard.edu/abs/2013A\%26ARv..21...70B} {21, 70}

\bibitem[\protect\citeauthoryear{{Cesarsky}}{{Cesarsky}}{1975}]{Cesarsky-1975}
{Cesarsky} C.~J.,  1975, International Cosmic Ray Conference, \href
  {http://adsabs.harvard.edu/abs/1975ICRC....2..634C} {2}

\bibitem[\protect\citeauthoryear{{Cesarsky} \& {Volk}}{{Cesarsky} \&
  {Volk}}{1978}]{Cesarsky-1978}
{Cesarsky} C.~J.,  {Volk} H.~J.,  1978, \aap, \href
  {http://adsabs.harvard.edu/abs/1978A%26A....70..367C} {70, 367}

\bibitem[\protect\citeauthoryear{{Cox}}{{Cox}}{1998}]{1998LNP...506..121C}
{Cox} D.~P.,  1998, in {Breitschwerdt} D.,  {Freyberg} M.~J.,   {Truemper} J.,
  eds,  Lecture Notes in Physics, Berlin Springer Verlag Vol. 506, IAU Colloq.
  166: The Local Bubble and Beyond. pp 121--131, \mn@doi{10.1007/BFb0104706}

\bibitem[\protect\citeauthoryear{{Cummings} et~al.,}{{Cummings}
  et~al.}{2016}]{Cummings-2016}
{Cummings} A.~C.,  et~al., 2016, \mn@doi [\apj] {10.3847/0004-637X/831/1/18},
  \href {http://adsabs.harvard.edu/abs/2016ApJ...831...18C} {831, 18}

\bibitem[\protect\citeauthoryear{{Dalgarno}}{{Dalgarno}}{2006}]{Dalgarno-2006}
{Dalgarno} A.,  2006, \mn@doi [Proceedings of the National Academy of Science]
  {10.1073/pnas.0602117103}, \href
  {http://adsabs.harvard.edu/abs/2006PNAS..10312269D} {103, 12269}

\bibitem[\protect\citeauthoryear{{Drury} \& {Strong}}{{Drury} \&
  {Strong}}{2017}]{Drury-2017}
{Drury} L.~O.~.,  {Strong} A.~W.,  2017, \mn@doi [\aap]
  {10.1051/0004-6361/201629526}, \href
  {http://adsabs.harvard.edu/abs/2017A%26A...597A.117D} {597, A117}

\bibitem[\protect\citeauthoryear{{Gabici} \& {Montmerle}}{{Gabici} \&
  {Montmerle}}{2015}]{Gabici-2015}
{Gabici} S.,  {Montmerle} T.,  2015, in 34th International Cosmic Ray
  Conference (ICRC2015). p.~29

\bibitem[\protect\citeauthoryear{{Goldreich} \& {Sridhar}}{{Goldreich} \&
  {Sridhar}}{1995}]{Goldreich-1995}
{Goldreich} P.,  {Sridhar} S.,  1995, \mn@doi [\apj] {10.1086/175121}, \href
  {http://adsabs.harvard.edu/abs/1995ApJ...438..763G} {438, 763}

\bibitem[\protect\citeauthoryear{{Hayakawa}, {Nishimura}  \&
  {Takayanagi}}{{Hayakawa} et~al.}{1961}]{Hayakawa-1961}
{Hayakawa} S.,  {Nishimura} S.,   {Takayanagi} T.,  1961, \pasj, \href
  {http://adsabs.harvard.edu/abs/1961PASJ...13..184H} {13, 184}

\bibitem[\protect\citeauthoryear{{Heyer} \& {Dame}}{{Heyer} \&
  {Dame}}{2015}]{Heyer-2015}
{Heyer} M.,  {Dame} T.~M.,  2015, \mn@doi [\araa]
  {10.1146/annurev-astro-082214-122324}, \href
  {http://adsabs.harvard.edu/abs/2015ARA%26A..53..583H} {53, 583}

\bibitem[\protect\citeauthoryear{{Indriolo}, {Fields}  \& {McCall}}{{Indriolo}
  et~al.}{2009}]{Indriolo-2009}
{Indriolo} N.,  {Fields} B.~D.,   {McCall} B.~J.,  2009, \mn@doi [\apj]
  {10.1088/0004-637X/694/1/257}, \href
  {http://adsabs.harvard.edu/abs/2009ApJ...694..257I} {694, 257}

\bibitem[\protect\citeauthoryear{{Ivlev}, {Dogiel}, {Chernyshov}, {Caselli},
  {Ko}  \& {Cheng}}{{Ivlev} et~al.}{2018}]{Ivlev-2018}
{Ivlev} A.~V.,  {Dogiel} V.~A.,  {Chernyshov} D.~O.,  {Caselli} P.,  {Ko}
  C.-M.,   {Cheng} K.~S.,  2018, \mn@doi [\apj] {10.3847/1538-4357/aaadb9},
  \href {http://adsabs.harvard.edu/abs/2018ApJ...855...23I} {855, 23}

\bibitem[\protect\citeauthoryear{{Jokipii}}{{Jokipii}}{2001}]{Jokipii-2001}
{Jokipii} R.,  2001, in {Ko} C.-M.,  ed.,  Astronomical Society of the Pacific
  Conference Series Vol. 241, The 7th Taipei Astrophysics Workshop on Cosmic
  Rays in the Universe. p.~223

\bibitem[\protect\citeauthoryear{{Krause}, {Morlino}  \& {Gabici}}{{Krause}
  et~al.}{2015}]{Krause-2015}
{Krause} J.,  {Morlino} G.,   {Gabici} S.,  2015, in 34th International Cosmic
  Ray Conference (ICRC2015). p.~518

\bibitem[\protect\citeauthoryear{{Krimigis}, {Decker}, {Roelof}, {Hill},
  {Armstrong}, {Gloeckler}, {Hamilton}  \& {Lanzerotti}}{{Krimigis}
  et~al.}{2013}]{Krimigis-2013}
{Krimigis} S.~M.,  {Decker} R.~B.,  {Roelof} E.~C.,  {Hill} M.~E.,  {Armstrong}
  T.~P.,  {Gloeckler} G.,  {Hamilton} D.~C.,   {Lanzerotti} L.~J.,  2013,
  \mn@doi [Science] {10.1126/science.1235721}, \href
  {http://adsabs.harvard.edu/abs/2013Sci...341..144K} {341, 144}

\bibitem[\protect\citeauthoryear{{Krolik} \& {Kallman}}{{Krolik} \&
  {Kallman}}{1983}]{Krolik-1983}
{Krolik} J.~H.,  {Kallman} T.~R.,  1983, \mn@doi [\apj] {10.1086/160897}, \href
  {http://adsabs.harvard.edu/abs/1983ApJ...267..610K} {267, 610}

\bibitem[\protect\citeauthoryear{{McKee}}{{McKee}}{1989}]{McKee-1989}
{McKee} C.~F.,  1989, \mn@doi [\apj] {10.1086/167950}, \href
  {http://adsabs.harvard.edu/abs/1989ApJ...345..782M} {345, 782}

\bibitem[\protect\citeauthoryear{{Meneguzzi}, {Audouze}  \&
  {Reeves}}{{Meneguzzi} et~al.}{1971}]{Meneguzzi-1971}
{Meneguzzi} M.,  {Audouze} J.,   {Reeves} H.,  1971, \aap, \href
  {http://adsabs.harvard.edu/abs/1971A%26A....15..337M} {15, 337}

\bibitem[\protect\citeauthoryear{{Morfill}}{{Morfill}}{1982}]{Morfill-1982}
{Morfill} G.~E.,  1982, \mn@doi [\apj] {10.1086/160470}, \href
  {http://adsabs.harvard.edu/abs/1982ApJ...262..749M} {262, 749}

\bibitem[\protect\citeauthoryear{{Morlino} \& {Gabici}}{{Morlino} \&
  {Gabici}}{2015}]{Morlino-2015}
{Morlino} G.,  {Gabici} S.,  2015, \mn@doi [\mnras] {10.1093/mnrasl/slv074},
  \href {http://adsabs.harvard.edu/abs/2015MNRAS.451L.100M} {451, L100}

\bibitem[\protect\citeauthoryear{{Nobukawa} et~al.,}{{Nobukawa}
  et~al.}{2015}]{Nobukawa-2015}
{Nobukawa} K.~K.,  et~al., 2015, \mn@doi [\apjl] {10.1088/2041-8205/807/1/L10},
  \href {http://ads.nao.ac.jp/abs/2015ApJ...807L..10N} {807, L10}

\bibitem[\protect\citeauthoryear{{Nobukawa} et~al.,}{{Nobukawa}
  et~al.}{2018}]{Nobukawa-2018}
{Nobukawa} K.~K.,  et~al., 2018, \mn@doi [\apj] {10.3847/1538-4357/aaa8dc},
  \href {http://ads.nao.ac.jp/abs/2018ApJ...854...87N} {854, 87}

\bibitem[\protect\citeauthoryear{{Osborne} \& {Ptuskin}}{{Osborne} \&
  {Ptuskin}}{1988}]{Osborne-1998}
{Osborne} J.~L.,  {Ptuskin} V.~S.,  1988, Soviet Astronomy Letters, \href
  {http://adsabs.harvard.edu/abs/1988SvAL...14..132O} {14, 132}

\bibitem[\protect\citeauthoryear{{Osterbrock} \& {Bochkarev}}{{Osterbrock} \&
  {Bochkarev}}{1989}]{Osterbrock-1989}
{Osterbrock} D.~E.,  {Bochkarev} N.~G.,  1989, \sovast, \href
  {http://adsabs.harvard.edu/abs/1989SvA....33..694O} {33, 694}

\bibitem[\protect\citeauthoryear{{Padovani} \& {Galli}}{{Padovani} \&
  {Galli}}{2013}]{Padovani-2013-rev}
{Padovani} M.,  {Galli} D.,  2013, in {Torres} D.~F.,  {Reimer} O.,  eds,
  Astrophysics and Space Science Proceedings Vol. 34, Cosmic Rays in
  Star-Forming Environments. p.~61 (\mn@eprint {arXiv} {1305.5393}),
  \mn@doi{10.1007/978-3-642-35410-6_6}

\bibitem[\protect\citeauthoryear{{Padovani}, {Galli}  \&
  {Glassgold}}{{Padovani} et~al.}{2009}]{Padovani-2009}
{Padovani} M.,  {Galli} D.,   {Glassgold} A.~E.,  2009, \mn@doi [\aap]
  {10.1051/0004-6361/200911794}, \href
  {http://adsabs.harvard.edu/abs/2009A%26A...501..619P} {501, 619}

\bibitem[\protect\citeauthoryear{{Padovani}, {Hennebelle}, {Marcowith}  \&
  {Ferri{\`e}re}}{{Padovani} et~al.}{2015}]{Padovani-2015}
{Padovani} M.,  {Hennebelle} P.,  {Marcowith} A.,   {Ferri{\`e}re} K.,  2015,
  \mn@doi [\aap] {10.1051/0004-6361/201526874}, \href
  {http://adsabs.harvard.edu/abs/2015A%26A...582L..13P} {582, L13}

\bibitem[\protect\citeauthoryear{{Padovani}, {Marcowith}, {Hennebelle}  \&
  {Ferri{\`e}re}}{{Padovani} et~al.}{2016}]{Padovani-2016}
{Padovani} M.,  {Marcowith} A.,  {Hennebelle} P.,   {Ferri{\`e}re} K.,  2016,
  \mn@doi [\aap] {10.1051/0004-6361/201628221}, \href
  {http://adsabs.harvard.edu/abs/2016A%26A...590A...8P} {590, A8}

\bibitem[\protect\citeauthoryear{{Phan}, {Morlino}  \& {Gabici}}{{Phan}
  et~al.}{2018}]{Phan-2018}
{Phan} V.~H.~M.,  {Morlino} G.,   {Gabici} S.,  2018, \mn@doi [\mnras]
  {10.1093/mnras/sty2235}, \href
  {http://adsabs.harvard.edu/abs/2018MNRAS.480.5167P} {480, 5167}

\bibitem[\protect\citeauthoryear{{Schlickeiser}, {Caglar}  \&
  {Lazarian}}{{Schlickeiser} et~al.}{2016}]{Schlickeiser-2016}
{Schlickeiser} R.,  {Caglar} M.,   {Lazarian} A.,  2016, \mn@doi [\apj]
  {10.3847/0004-637X/824/2/89}, \href
  {http://adsabs.harvard.edu/abs/2016ApJ...824...89S} {824, 89}

\bibitem[\protect\citeauthoryear{{Silk} \& {Norman}}{{Silk} \&
  {Norman}}{1983}]{Silk-1983}
{Silk} J.,  {Norman} C.,  1983, \mn@doi [\apjl] {10.1086/184115}, \href
  {http://adsabs.harvard.edu/abs/1983ApJ...272L..49S} {272, L49}

\bibitem[\protect\citeauthoryear{{Skilling} \& {Strong}}{{Skilling} \&
  {Strong}}{1976}]{Skilling-1976}
{Skilling} J.,  {Strong} A.~W.,  1976, \aap, \href
  {http://adsabs.harvard.edu/abs/1976A%26A....53..253S} {53, 253}

\bibitem[\protect\citeauthoryear{{Sridhar} \& {Goldreich}}{{Sridhar} \&
  {Goldreich}}{1994}]{Sridhar-1994}
{Sridhar} S.,  {Goldreich} P.,  1994, \mn@doi [\apj] {10.1086/174600}, \href
  {http://adsabs.harvard.edu/abs/1994ApJ...432..612S} {432, 612}

\bibitem[\protect\citeauthoryear{{Stone}, {Cummings}, {McDonald}, {Heikkila},
  {Lal}  \& {Webber}}{{Stone} et~al.}{2013}]{Stone-2013}
{Stone} E.~C.,  {Cummings} A.~C.,  {McDonald} F.~B.,  {Heikkila} B.~C.,  {Lal}
  N.,   {Webber} W.~R.,  2013, \mn@doi [Science] {10.1126/science.1236408},
  \href {http://adsabs.harvard.edu/abs/2013Sci...341..150S} {341, 150}

\bibitem[\protect\citeauthoryear{{Strong}, {Porter}, {Digel},
  {J{\'o}hannesson}, {Martin}, {Moskalenko}, {Murphy}  \& {Orlando}}{{Strong}
  et~al.}{2010}]{Strong-2010}
{Strong} A.~W.,  {Porter} T.~A.,  {Digel} S.~W.,  {J{\'o}hannesson} G.,
  {Martin} P.,  {Moskalenko} I.~V.,  {Murphy} E.~J.,   {Orlando} E.,  2010,
  \mn@doi [\apjl] {10.1088/2041-8205/722/1/L58}, \href
  {http://adsabs.harvard.edu/abs/2010ApJ...722L..58S} {722, L58}

\bibitem[\protect\citeauthoryear{{Tatischeff} \& {Gabici}}{{Tatischeff} \&
  {Gabici}}{2018}]{Tatischeff-2018-review}
{Tatischeff} V.,  {Gabici} S.,  2018, \mn@doi [Annual Review of Nuclear and
  Particle Science] {10.1146/annurev-nucl-101917-021151}, \href
  {http://adsabs.harvard.edu/abs/2018ARNPS..68..377T} {68, 377}

\bibitem[\protect\citeauthoryear{{Thornbury} \& {Drury}}{{Thornbury} \&
  {Drury}}{2014}]{Drury-2014}
{Thornbury} A.,  {Drury} L.~O.,  2014, \mn@doi [\mnras]
  {10.1093/mnras/stu1080}, \href
  {http://adsabs.harvard.edu/abs/2014MNRAS.442.3010T} {442, 3010}

\bibitem[\protect\citeauthoryear{{Tomasko} \& {Spitzer}}{{Tomasko} \&
  {Spitzer}}{1968}]{Spitzer-1968}
{Tomasko} M.~G.,  {Spitzer} L.,  1968, The Astronomical Journal Supplement,
  \href {http://adsabs.harvard.edu/abs/1968AJS....73S..37T} {73, 37}

\bibitem[\protect\citeauthoryear{{Trotta}, {J{\'o}hannesson}, {Moskalenko},
  {Porter}, {Ruiz de Austri}  \& {Strong}}{{Trotta}
  et~al.}{2011}]{Galprop-2011-propa}
{Trotta} R.,  {J{\'o}hannesson} G.,  {Moskalenko} I.~V.,  {Porter} T.~A.,
  {Ruiz de Austri} R.,   {Strong} A.~W.,  2011, \mn@doi [\apj]
  {10.1088/0004-637X/729/2/106}, \href
  {http://adsabs.harvard.edu/abs/2011ApJ...729..106T} {729, 106}

\makeatother
\end{thebibliography}


\bsp	
\label{lastpage}
\end{document}